\documentclass[12pt]{article}
\usepackage{a4wide}
\usepackage{epsfig}
\usepackage{amsmath}

\newcommand{\half}{{\textstyle\frac{1}{2}}}

\newlength{\absize}
\catcode`@=11
\def\citer{\@ifnextchar
[{\@tempswatrue\@citexr}{\@tempswafalse\@citexr[]}}

\def\@citexr[#1]#2{\if@filesw\immediate
  \write\@auxout{\string\citation{#2}}\fi
  \def\@citea{}\@cite{\@for\@citeb:=#2\do
    {\@citea\def\@citea{--\penalty\@m}\@ifundefined
       {b@\@citeb}{{\bf ?}\@warning
       {Citation `\@citeb' on page \thepage \space undefined}}%
\hbox{\csname b@\@citeb\endcsname}}}{#1}}
\catcode`@=12

\begin{document}
  \thispagestyle{empty}
  \pagestyle{empty}
  \renewcommand{\thefootnote}{\fnsymbol{footnote}}
\newpage\normalsize
    \pagestyle{plain}
    \setlength{\baselineskip}{4ex}\par
    \setcounter{footnote}{0}
    \renewcommand{\thefootnote}{\arabic{footnote}}
\newcommand{\preprint}[1]{%
  \begin{flushright}
    \setlength{\baselineskip}{3ex} #1
  \end{flushright}}
\renewcommand{\title}[1]{%
  \begin{center}
    \LARGE #1
  \end{center}\par}
\renewcommand{\author}[1]{%
  \vspace{2ex}
  {\Large
   \begin{center}
     \setlength{\baselineskip}{3ex} #1 \par
   \end{center}}}
\renewcommand{\thanks}[1]{\footnote{#1}}

\begin{center}
{\large \bf Structures of $q$-Deformed Currents}
\end{center}
\vspace{0.5cm}
\begin{center}
Jian-zu Zhang$^{a,b, \S}$
\end{center}
\vspace{0.5cm}
\begin{center}
$^a$ Institute for Theoretical Physics, Box 316, East China
University of Science and Technology, Shanghai 200237, P. R. China\\
$^b$  Department of Physics, University of Kaiserslautern, PO Box
3049, D-67653  Kaiserslautern, Germany
\end{center}
\vspace{0.5cm}
\begin{abstract}
The non-perturbation and perturbation structures of
the $q$-deformed probability currents are studied.
According to two ways of realizing the $q$-deformed Heisenberg algebra
by the undeformed operators,
the perturbation  structures of two $q$-deformed probability currents
 are explored in detail.
Locally the structures of two  perturbation $q$-deformed  probability
 currents are different,
one is explicitly potential dependent; the other is  not.
But their total contributions  to the whole space are the same.
\end{abstract}
\begin{flushleft}
PACS: 03. 65. -w \qquad   03. 65. ca
${^\S}$ E-mail address:  jzzhang@ecust.edu.cn\\
\hspace{3.1cm} jzzhang@physik.uni-kl.de
\end{flushleft}
\clearpage
Recently  the $q$-deformed quantum theory, as a possible
modification of the ordinary quantum theory at extremely small
space scale, say, much smaller than  $10^{-18}$ cm, has obtained
attention. In literature different frameworks of $q$-deformed
quantum theories were established \citer{Schwenk,JZZ02}. In order
to establish a consistent framework in $q$-deformed quantum
theories three delicate points must be considered: the
construction of the simultaneous hermitian position and momentum
operators; the establishment of the correspondence of the position
and momentum operators to the $q$-deformed annihilation and
creation operators; and the reduction of the $q$-deformed
annihilation and creation operators to the undeformed ones. In the
framework of the $q$-deformed Heisenberg algebra developed in
Refs.~\cite{Hebecker,Fichtmuller} the above three aspects are
investigated in detail. This framework is self-consistent. New
features, both in the uncertainty relations and dynamics,
 in this framework are explored.
The $q$-deformed uncertainty relation essentially deviates from
the Heisenberg one \cite{JZZ98,JZZ99,OZ00,JZZ02}: Heisenberg's
minimal uncertainty relation is undercut. In a special
$q$-deformed squeezed states  a new critical phenomenon  is
explored\cite{OZ00}: at a critical point the variance of one
component of a quadrature of light field approaches zero, but the
variance of the conjugate component remains finite. Such critical
phenomenon  is forbidden by Heisenberg's uncertainty relations,
but allowed by the $q$-deformed uncertainty relations. In dynamics
the non-perturbation energy spectrum of the $q$-deformed
Schr\"odinger equation exhibits an exponential structure
\cite{LW,Fichtmuller,JZZ00} with new degrees of freedom and new
quantum numbers.
 Using such  an exponential structure the spectrum of quark-lepton is
 explained \cite{JZZ00}.
In the perturbation aspects the $q$-deformed dynamics also
exhibits a new feature: the perturbation expansion of the
$q$-deformed Hamiltonian possesses complicated structures, which
amount to additional momentum-dependent interactions
\cite{Hebecker,Fichtmuller,JZZ98,JZZ00,ZO01,JZZ01,JZZ02}.
 Furthermore, corresponding to two ways of realizing the $q$-deformed
operators by the undeformed ones there are two $q$-perturbation
momentum-dependent Hamiltonians,
one originates from the perturbation expansion of the potential in one
configuration  space,
the other  originates from the perturbation expansion of the kinetic energy
 in the other configuration  space.
At the level of operators, they are different. But they contribute
the same shifts to the undeformed energy spectrum.
\cite{ZO01,JZZ01,JZZ02}.

In this paper the non-perturbation and perturbation structures of
the $q$-deformed probability currents are investigated.
The study of the perturbation $q$-deformed currents  explores that
the essential deviation of the $q$-deformed quantum mechanics from
the ordinary quantum mechanics is in their {\it local} structure.
The perturbation  structures of the $q$-deformed probability currents
in  these two $q$-perturbation approaches are investigated in detail.
 In one approach the components of the $q$-deformed  perturbation
probability current possess different ranks, describing space
derivatives of different order of the corresponding sub-current of
different levels. Locally the structures of two perturbation
$q$-deformed   probability
 currents are different,
one is explicitly potential dependent;  the other is not.
But their total contributions  to the whole space are the same.

In the following we first review the background of the $q$-deformed dynamics.
In terms of the $q$-deformed phase space variables ---
position and momentum operators $X$ and $P$
the following $q$-deformed Heisenberg algebra has been
developed \cite{Hebecker, Fichtmuller}:
\begin{equation}
\label{Eq:q-algebra}
q^{1/2}XP-q^{-1/2}PX=iU, \qquad
UX=q^{-1}XU, \qquad
UP=qPU,
\end{equation}
where $X$ and $P$ are hermitian and $U$ is unitary:
$X^{\dagger}=X$, $P^{\dagger}=P$, $U^{\dagger}=U^{-1}$.
Compared to the Heisenberg algebra the operator $U$ is a new member,
called scaling operator.
Necessity of introducing the operator $U$ is as follows.

Simultaneous hermitian of  $X$ and  $P$ is a delicate point in
the $q$-deformed dynamics.
Definition of algebra
(\ref{Eq:q-algebra}) is based on definition of the hermitian momentum
operator $P$.
However, if $X$ is assumed to be a hermitian operator in a Hilbert space
the $q$-deformed derivative \cite{Fichtmuller,Wess}
\begin{equation}
\label{Eq: q-derivative}
\partial_X X=1+qX\partial_X,
\end{equation}
which codes non-commutativity of space, shows that
 the usual quantization rule $P\to -i\partial_X$ does not yield a
hermitian
momentum operator. A hermitian momentum operator $P$ is related to
$\partial_X$ and $X$ in a nonlinear way by introducing a scaling operator
$U$ \cite{Fichtmuller}
\begin{equation}
\label{Eq:scaling}
U^{-1}\equiv q^{1/2}[1+(q-1)X\partial_X], \qquad
\bar\partial_X\equiv -q^{-1/2}U\partial_X, \qquad
P\equiv -\frac{i}{2}(\partial_X-\bar\partial_X),
\end{equation}
where $\bar\partial_X$ is the conjugate of $\partial_X$.
The operator $U$ is introduced in the definition of the hermitian
momentum,
thus it closely relates to properties of dynamics and plays an
essential role in the $q$-deformed quantum mechanics.
Nontrivial properties of  $U$ imply that the algebra (\ref{Eq:q-algebra})
has a richer structure than the Heisenberg commutation relation.
In Eq.~(\ref{Eq:q-algebra}) the parameter $q$ is a fixed real number.
It is important to make distinctions for different realizations of
the $q$-algebra by different ranges of $q$ values \citer{Zachos,Solomon}.
Following Refs.~\cite{Hebecker,Fichtmuller}  we only consider the case
$q>1$
 in this paper. The reason is that such choice of  the parameter $q$
leads to consistent dynamics.
In the limit $q\to 1^+$ the scaling operator $U$ reduces
to a unit operator, thus the algebra (\ref{Eq:q-algebra}) reduces to
the Heisenberg commutation relation.

Such defined hermitian momentum $P$ leads to the $q$-deformation effects,
which are exhibited by the dynamical equation.
Eq. (\ref{Eq:scaling}) shows that the  momentum $P$ depends  non-linearly
on $X$ and $\partial_X$.
Thus the $q$-deformed
Schr\"odinger equation is difficult to treat.
The perturbation treatment of the $q$-deformed Schr\"odinger equation
is based on realizing the $q$-deformed operators  $X$, $P$ and  $U$
by undeformed variables.
There are two pairs of  undeformed variables \cite{Fichtmuller}.

(I) Variables $\hat x$, $\hat p$ of the ordinary quantum mechanics,
where $\hat x$, $\hat p$ satisfy:
$[ \hat x, \hat p ]=i$, $\hat x=\hat x^{\dagger}$,
 $\hat p=\hat p^ {\dagger}$.
The $q$-deformed operators $X$, $P$ and $U$ are related to $\hat x$,
$\hat p$ by:
\begin{equation}
\label{Eq:P-p}
X= \frac{[\hat z+\half]}{\hat z+\half}\hat x,  \qquad
P=\hat p, \qquad
U= q^{\hat z},
\end{equation}
where $\hat z=-\frac{i}{2}(\hat x\hat p+\hat p\hat x)$ and
$[A]$ is the $q$-deformation of $A$, defined by
$[A]=(q^A-q^{-A})/(q-q^{-1})$.
It is easy to check that such defined $X$, $P$ and $U$ satisfy
Eq.~(\ref{Eq:q-algebra}).

(II) Variables $\tilde x$ and $\tilde p$ of an undeformed algebra,
which are obtained by a  transformation
of $\hat x$ and $\hat p$:
\begin{equation}
\label{Eq:tilde}
\tilde x=\hat x F^{-1}(\hat z), \qquad \tilde p= F(\hat z)\hat p,
\end{equation}
where
\begin{equation}
\label{Eq:F(z)}
F^{-1}(\hat z)= \frac{[\hat z-\half]}{\hat z-\half},
\end{equation}
 Such defined variables $\tilde x$ and $\tilde p$ also satisfy
undeformed algebra: $[ \tilde x, \tilde p ]=i$, and $\tilde
x=\tilde x^{\dagger}$,\quad$\tilde p=\tilde p^{ \dagger}$. Thus
$\tilde p=-i\partial_{\tilde x}$. The operator $F^{-1}(\hat z)$ is
non-unitary: Using $[A]^{\dagger}=[A^{\dagger}],$ $[-A]=-[A],$ and
$\hat z^{\dagger}=-\hat z,$ we have
\begin{equation*}
\nonumber
(F^{-1}(\hat z))^{\dagger}=F^{-1}(-\hat z)
=\frac{[\hat z+\half]}{\hat z+\half}, \quad
F^{-1}(\hat z)(F^{-1}(\hat z))^{\dagger}\ne I.
\end{equation*}
Though the transformation $F^{-1}(\hat z)$ maintains the
commutation relation $[\hat x, \hat p],$ the essential point is
that it does {\it not} maintain the inner product $\langle
\psi|\phi \rangle.$ It is not clear whether $F^{-1}(\hat z)$ leads
to the same physical consequences in the $(\hat x,\hat p)$ system
and the $(\tilde x,\tilde p)$ system. A detailed study of
perturbation aspects in these two systems is necessary. The
$q$-deformed operators $X$, $P$ and $U$ are related to  $\tilde
x$ and $\tilde p$ as follows:
\begin{equation}
\label{Eq:X-x}
X=\tilde x, \qquad P=F^{-1}(\tilde z) \tilde p, \qquad
U= q^{\tilde z},  \qquad
\tilde z=-\frac{i}{2}(\tilde x\tilde p + \tilde p\tilde x)
\end{equation}
with $F^{-1}(\tilde z)$ defined by Eq.~(\ref{Eq:F(z)}) for variables
($\tilde  x$, $\tilde p$).
From Eqs.~(\ref{Eq:tilde})--(\ref{Eq:X-x}) it follows that $X$,
$P$ and $U$ satisfy Eq.~(\ref{Eq:q-algebra}).
 and Eq. (\ref{Eq:X-x})  is equivalent to Eq. (\ref{Eq:P-p}).

Now we first consider the non-perturbation probability
current of the $q$-deformed dynamics.

The $q$-deformed phase space ($X$, $P$) governed by the
$q$-algebra (\ref{Eq:q-algebra}) is a $q$-deformation of the phase
space ($\hat x$, $\hat p$) of the ordinary quantum mechanics, thus
all the machinery of the ordinary quantum mechanics can be applied
to the $q$-deformed quantum mechanics. It means that dynamical
equations of a quantum system are the same for the undeformed
operators and for the $q$-deformed operators ($X$, $P$). That is,
the time-dependent Schr\"odinger equation with the $q$-deformed
Hamiltonian  $H(X,P)=\frac{1}{2\mu}P^{2}+V(X)$ is $i \partial_t
\psi(X,t)=H(X,P)\psi(X,t).$ Using Eqs.~(\ref{Eq: q-derivative})
and (\ref{Eq:scaling}), we rewrite $P^2$ as
\begin{equation}
\label{Eq:PP} P^2= -\frac{1}{4}\partial_X \bigl(1+\eta
U+U^2\bigl)\partial_X,
\end{equation}
where $\eta=(q^{1/2}+q^{-1/2}),$ then it follows that the
$q$-deformed continuity equation of the  probability conservation
reads
$$\partial_t \rho (X, t)+\partial_X j_q(X, t)=S_q(X,t),$$
where the position probability density $\rho (X,
t)=\psi^{*}(X,t)\psi(X,t);$ the q-deformed probability current
density is
\begin{eqnarray}
\label{Eq:jq}
j_q(X,t)= -\frac{i}{8\mu}\Bigl\{\bigl[(1+\eta U
+U^2)\bigl(\partial_X \psi(X,t)\bigr)_q\bigr]\psi^{*}(X,t) \nonumber \\
-\bigl[(1+\eta U+U^2)\bigl(\partial_X\psi(X,t)\bigr)_q\bigr]^{*}\psi(X,t)
\Bigr\}.
\end{eqnarray}
Because of Eq.~(\ref{Eq:PP}) the second term in Eq.~(\ref{Eq:jq})
is written as \\
$\bigl[(1+\eta
U+U^2)\bigl(\partial_X\psi(X,t)\bigr)_q\bigr]^{*}\psi(X,t),$ not as \\
$\bigl(1+\eta
U+U^2\bigr)\bigl(\partial_X\psi(X,t)\bigr)_q^{*}\psi(X,t).$ In the
above $\bigl(\partial_X \psi(X,t)\bigr)_q$ is the q-deformed
partial derivative with the bracket.
\footnote{From Eq.~(\ref{Eq: q-derivative}) of the q-derivative it follows
that
\begin{equation*}
\partial_X X^n=q^n X^n \partial_X+\bigl(\partial_X X^n\bigr)_q, \nonumber
\end{equation*}
where the q-deformed partial derivative $\bigl(\partial_X X^n\bigr)_q$
 with the bracket
 is defined as
\begin{equation*}
\bigl(\partial_X X^n\bigr)_q\equiv \sum_{k=0}^{n-1}q^k X^{n-1}
=\frac{q^n-1}{q-1}X^{n-1}. \nonumber
\end{equation*}
The q-deformed partial derivative for function $f(x),$ which has well
defined Tailor expansion, reads
\begin{equation*}
\partial_X f(X)=f(qX)\partial_X +\bigl(\partial_X f(X)\bigr)_q, \nonumber
\end{equation*}
Notice that in the first term the variable of the function $f$ is  $qX.$
where the  q-deformed partial derivative $\bigl(\partial_X f(X)\bigr)_q$
 with the bracket is  defined as
\begin{equation*}
\bigl(\partial_X f(X)\bigr)_q\equiv \sum_{k=1}^\infty \frac{f^{n}(0)}{n!}
\bigl(\partial_X X^n\bigr)_q. \nonumber
\end{equation*}
When $q\to 1^+$ the  q-deformed partial derivative with the
bracket reduces to undeformed one, $\bigl(\partial_X
X^n\bigr)_q\to n X^{n-1}$ and $\bigl(\partial_X f(X)\bigr)_q \to
f^{\prime}(X).$ The q-deformed Leibniz rule  for $f(X)g(X)$ is
\begin{equation*}
\partial_X \bigl(f(X)g(X)\bigr)=\bigl(\partial_X f(X)\bigr)_qg(X)
+f(qX)\bigl(\partial_X g(X)\bigr)_q. \nonumber
\end{equation*}
Notice again that in the second term the variable of the function $f$ is
 $qX.$
This leads to a involved structure of the q-deformed current.}
When $q\to 1^+$, the scaling operator $U$ reduces to the unit
operator, thus the q-deformed current term $j_q$ reduces to the
ordinary probability current density $j_{un}(X,t)=
-\frac{i}{2\mu}\{ \psi^{*}(X,t)\partial_X \psi(X,t)
-[\partial_X\psi^{*}(X,t)]\psi(X,t)\}.$

In the above $S_q$ is the q-deformed source term,
\begin{eqnarray}
\label{Eq:Sq}
S_q(X,t)=-\frac{i}{8\mu}\Bigl\{\bigl[(1+\eta U
+U^2)\bigl(\partial_X \psi(qX,t)\bigr)_q\bigr]
\bigl(\partial_X\psi^{*}(X,t)\bigr)_q
 \nonumber \\
-\bigl[(1+\eta U+U^2)\bigl(\partial_X\psi(qX,t)\bigr)_q\bigr]^{*}
\bigl(\partial_X\psi(X,t)\bigr)_q\Bigr\}.
\end{eqnarray}
Similar to Eq.~(\ref{Eq:jq})
the second term in Eq.~(\ref{Eq:Sq}) is written as
\\
$\bigl[(1+\eta
U+U^2)\bigl(\partial_X\psi(qX,t)\bigr)_q\bigr]^{*}
\bigl(\partial_X\psi(X,t)\bigr)_q,$ not as \\
$\bigl(1+\eta U+U^2\bigr)\bigl(\partial_X\psi(qX,t)\bigr)_q^{*}
\bigl(\partial_X\psi(X,t)\bigr)_q.$ The local conservation of the
probability requires that the source term $S_q$ should vanish.
This inference can be drawn from the following considerations. It
is noticed that when $q\to1^+,$ in Eq.~(\ref{Eq:Sq}) the first
term $\bigl[(1+\eta U+U^2)\bigl(\partial_X
\psi(qX,t)\bigr)_q\bigr] \bigl(\partial_X\psi^{*}(X,t)\bigr)_q$
reduces to a real quantity $4\bigl(\partial_X \psi(X,t)\bigr)
\bigl(\partial_X\psi^{*}(X,t)\bigr).$ The same conclusion holds
for the second term in Eq.~(\ref{Eq:Sq}). The $q$ dependent
operator $U$ is introduced in the definition of the hermitian
momentum $P.$ For any value of the parameter $q$ the momentum $P$
should maintain the same physical property. This requires that for
any value of the parameter $q$ the operator $U$ should be
singularity free and maintain the same analytical property.
Therefore a continuous variation of the parameter $q$ does not
change the real properties of the two terms in Eq.~(\ref{Eq:Sq}),
the two terms exactly cancel each other which leads to $S_q=0.$

Now we  discuss the  perturbation $q$-deformed probability
currents.

If the $q$-deformed quantum mechanics is a realistic physical theory
at  short distances much smaller than $10^{-18}$~cm,
its correction to the ordinary quantum mechanics must be extremely small
 in the energy range of nowadays experiments.
This means that the parameter $q$ must be extremely close to one,
the  perturbation investigation of the $q$-deformed dynamics is
meaningful. So we can let $q=e^{f}=1+f+f^2$ with $0<f\ll1$ and it
is accurate enough in the perturbation expansion to the order
$f^2.$

In the  ($\hat x$, $\hat p$) system from Eq.~(\ref{Eq:P-p}) it follows
that
 $X$ is represented as a  non-linear function of ($\hat x$, $\hat p$):
\begin{equation}
\label{Eq:X-variable}
X=i(q-q^{-1})^{-1} \bigl( q^{(\hat z+1/2)} -
q^{-(\hat z+1/2)}\bigr)\hat p^{-1}.
\end{equation}
Using Eq.~(\ref{Eq:X-variable}), to the order $f^2,$ the
perturbation expansion of $X$ reads
\begin{equation}
\label{Eq:X-perturbation}
X=\hat x  + f^2 g(\hat x, \hat p), \qquad
g(\hat x,\hat p)=-\frac{1}{6}(1+
\hat x  \hat p  \hat x  \hat p )\hat x.
\end{equation}
For regular potentials $V(X)$ which are singularity free, to the
order $f^2$ in the perturbation expansion, such potentials can be
expressed by the undeformed variables ($\hat x$, $\hat p$) as
\begin{equation}
\label{Eq:q-potential}
V(X)=V(\hat x) +\hat H^{(q)}_I(\hat x,\hat p),
\end{equation}
where the perturbation Hamiltonian is
\begin{equation}
\label{Eq:H-q-summed}
\hat H^{(q)}_I(\hat x,\hat p)
=\frac{f^2}{6}
\bigl\{ \hat x^3 V'(\hat x) \partial^2_{\hat x}
+ \bigl[\hat x^3 V^{''}(\hat x) +3 \hat x^2 V^{'}(\hat x)\bigr]
\partial_{\hat x}
+{\textstyle\frac{1}{3}}\hat x^3 V^{'''}(\hat x)
+{\textstyle\frac{3}{2}}\hat x^2 V^{''}(\hat x)\bigr\}.
\end{equation}

In the  ($\tilde x$, $\tilde p$) system using Eq.~(\ref{Eq:X-x})
the perturbation expansions of the momentum $P$ and the kinetic energy
$P^{2}/(2\mu),$ to the order $f^2,$ read
\begin{equation}
\label{Eq:P-perturbation}
P=\tilde p  + f^2 h(\tilde x ,\tilde p),\qquad
h(\tilde  x,\tilde p)
=-\frac{1}{6}(1+
\tilde  p\tilde x\tilde  p\tilde x )\tilde p,
\end{equation}
\begin{equation}
\label{Eq:Ek-q}
\frac{1}{2\mu}P^{2}=\frac{1}{2\mu}\tilde p^{2}+
\tilde H^{(q)}_I(\tilde x,\tilde p),
\end{equation}
where the  perturbation Hamiltonian is
\begin{eqnarray}
\label{Eq:tilde-Hq}
\tilde H^{(q)}_I(\tilde x,\tilde p)&=& \frac{1}{2\mu}f^2
\bigl[\tilde p\, h(\tilde  x,\tilde p)
+h(\tilde  x,\tilde p)\, \tilde p\bigr]
\nonumber\\
&=& -\frac{1}{12\mu}f^2 \bigl[ 2\tilde x^2 \partial_{\tilde x}^4+
8\tilde x  \partial_{\tilde x}^3+
3\partial_{\tilde x}^2 \bigr]
\end{eqnarray}
From Eqs.~(\ref{Eq:q-potential}),  (\ref{Eq:H-q-summed}), (\ref{Eq:Ek-q})
and
(\ref{Eq:tilde-Hq}),
it follows that the perturbation expansion of the $q$-deformed Hamiltonian
$H(X,P)$  can be written down
in the  $(\hat x,\hat p)$ system
or the $(\tilde  x,\tilde p)$ system.
In  the  $(\hat x,\hat p)$ system we have
\begin{equation}
\label{Eq:hat-H}
H(X(\hat x,\hat p),P(\hat x,\hat p))=H_{\rm un}(\hat x,\hat p)+
\hat H^{(q)}_I (\hat x,\hat p);
\end{equation}
in the  $(\tilde  x,\tilde p)$ system we have
\begin{equation}
\label{Eq:tilde-H}
H(X(\tilde x,\tilde p),P(\tilde x,\tilde p))=H_{\rm un}(\tilde x,\tilde p)
+\tilde H^{(q)}_I (\tilde x,\tilde p).
\end{equation}
In the above
\begin{equation}
\label{Eq:xi-Hun}
H_{\rm un}(\xi,\kappa)=\frac{1}{2\mu}\kappa^{2}+V(\xi)
\end{equation}
is the corresponding undeformed Hamiltonian in the $(\xi,\kappa)$ system,
where $(\xi,\kappa)$ represents
$(\hat x,\hat p)$ or $(\tilde  x,\tilde p)$.

Eqs.~(\ref{Eq:H-q-summed}) and (\ref{Eq:tilde-Hq}) show that the above two
perturbation Hamiltonians $\hat H^{(q)}_I (\hat x,\hat p)$
and
$\tilde H^{(q)}_I (\tilde x,\tilde p)$ originate,  separately, from
the perturbation expansions  of the potential and the kinetic energy.
At the level of operator they are different.

In the ($\tilde x$, $\tilde p$) system, to the order $f^2,$ from
the $q$-deformed Schr\"odinger equation with
the Hamiltonians  (\ref{Eq:tilde-H}), (\ref{Eq:xi-Hun}) and
 (\ref{Eq:tilde-Hq})
we obtain
\begin{eqnarray}
\partial_t \left [ \psi^{*}(\tilde x,t)\psi(\tilde x,t) \right ]
&=& \psi^{*}(\tilde x,t)\left[H_{\rm un}(\tilde x,\tilde p)
+\tilde H^{(q)}_I (\tilde x,\tilde p)\right]\psi(\tilde x,t)
\nonumber\\
&& - \Big\{\left[H_{\rm un}(\tilde x,\tilde p)
+\tilde H^{(q)}_I (\tilde x,\tilde p)\right]\psi^{*}(\tilde x,t)\Big\}
\psi(\tilde x,t). \nonumber
\end{eqnarray}
 The terms with $H_{\rm un}(\tilde x,\tilde p)$
lead to the undeformed  current
$j_{\rm un}(\tilde x, t)= -\frac{i}{2\mu}\{ \psi^{*}(\tilde x,t)
\partial_{\tilde x} \psi(\tilde x,t)
-[\partial_{\tilde x}\psi^{*}(\tilde x,t)]\psi(\tilde x,t)\}$
in the $(\tilde x,\tilde p)$ system. The terms with $\tilde
H^{(q)}_I(\tilde x,\tilde p)$ lead to the perturbation
contribution of the $q$-deformed current. The perturbation
structure of the $q$-deformed current is involved.
 As an example, we show the perturbation structure
 contributed by the term  $\tilde x^2 \partial_{\tilde x}^4$
in Eq.~(\ref{Eq:tilde-Hq})  in detail.
\begin{eqnarray}
&-\frac{i}{12\mu} \bigl \{ \psi^{*}(\tilde x,t) 2\tilde x^2
\partial_{\tilde x}^4 \psi(\tilde x,t)-
2\bigl [\tilde x^2\partial_{\tilde x}^4\psi^{*}(\tilde x,t) \bigr
]\psi(\tilde x,t)
\bigr \} =
\nonumber \\
&\partial_{\tilde x}\bigl[ 2j_{q2}(\tilde x, t)
-12j_{q1}(\tilde x, t)
-4j_{q0}(\tilde x, t)
+2j_{\rm un}(\tilde x, t)\bigr]
+8\tilde s^{(2)}_{q1}(\tilde x, t),\nonumber
\end{eqnarray}
where
\begin{equation}
\label{Eq:jq0}
j_{q0}(\tilde x, t)=\mathcal{J}_{q0}= -\frac{i}{12\mu}
\bigl \{\bigl [  \partial_{\tilde x} \psi^{*}(\tilde x,t) \bigr ]\tilde
x^2
\partial_{\tilde x}^2\psi(\tilde x,t)-
\bigl [ \partial_{\tilde x}^2\psi^{*}(\tilde x,t) \bigr ]\tilde x^2
\partial_{\tilde x} \psi(\tilde x,t)
\bigr \} .
\end{equation}
\begin{equation}
\label{Eq:jq1}
j_{q1}(\tilde x, t)=\partial_{\tilde x}\mathcal{J}_{q1}, \qquad
\mathcal{J}_{q1}= -\frac{i}{12\mu}
\bigl \{ \psi^{*}(\tilde x,t)\tilde x \partial_{\tilde x}\psi(\tilde x,t)-
\bigl [ \partial_{\tilde x}\psi^{*}(\tilde x,t) \bigr ]\tilde x
\psi(\tilde x,t)
\bigr \}.
\end{equation}
\begin{equation}
\label{Eq:jq2}
j_{q2}(\tilde x, t)=\partial_{\tilde x}^2\mathcal{J}_{q2}, \qquad
\mathcal{J}_{q2}= -\frac{i}{12\mu}
\bigl \{ \psi^{*}(\tilde x,t)\tilde x^2
\partial_{\tilde x}\psi(\tilde x,t)-
\bigl [ \partial_{\tilde x}\psi^{*}(\tilde x,t) \bigr ]\tilde x^2
 \psi(\tilde x,t)
\bigr \}.
\end{equation}
\begin{equation}
\label{Eq:sq}
\tilde s^{(2)}_{q1}(\tilde x, t)=-\frac{i}{12\mu}
\bigl \{\bigl [  \partial_{\tilde x} \psi^{*}(\tilde x,t) \bigr ]\tilde x
\partial_{\tilde x}^2\psi(\tilde x,t)-
\bigl [ \partial_{\tilde x}^2\psi^{*}(\tilde x,t) \bigr ]\tilde x
\partial_{\tilde x} \psi(\tilde x,t) \bigr \}.
\end{equation}
In the above $\tilde s^{(2)}_{q1}(\tilde x, t)$ is the
$q$-deformed probability source from the terms $\tilde x^2
\partial_{\tilde x}^4.$  Summing the total contributions
of $\tilde H^{(q)}_I(\tilde x,\tilde p),$ we find that the
contributions to $\tilde s^{(2)}_q$ from the terms $\tilde x^2
\partial_{\tilde x}^4$ and $\tilde x  \partial_{\tilde x}^3$
 in Eq.~(\ref{Eq:tilde-Hq})
 exactly cancel each other.
Thus the $q$-deformed continuity equation of the probability conservation is
$$\partial_t \rho (\tilde x, t)+\partial_{\tilde x}
\left[ j_{\rm un}(\tilde x, t) + f^2\tilde j^{(2)}_q(\tilde x, t)\right]
=0$$
where the perturbation $q$-deformed current
$\tilde j^{(2)}_q(\tilde x,t)$ is
\begin{equation}
\label{Eq:jq-tilde}
\tilde j^{(2)}_q(\tilde x, t)= 2j_{q2}(\tilde x, t)
-4j_{q1}(\tilde x, t)
-4j_{q0}(\tilde x, t)
-\frac{1}{6} j_{\rm un}(\tilde x, t).
\end{equation}
The perturbation $q$-deformed current $\tilde j^{(2)}_q$ possesses
different components $j_{qi}.$ A current is called the basic
current if it is not a space derivative of another current. The
currents $j_{un}, j_{q0}=\mathcal{J}_{q0}, \mathcal{J}_{q1},
\mathcal{J}_{q2}$ are the basic current. A current is called the
composed current with $i$'s sub-currents if it is space
derivatives of $i$ order of a basic current. The currents $j_{q1}$
is a composed current with one sub-current $\mathcal{J}_{q0}.$ The
composed current $j_{q2}$ includes two sub-currents: the current
$\mathcal{J}_{q2}$ and the effective sub-current
$\mathcal{J}_{q1}^{eff}\equiv
\partial_{\tilde x}\mathcal{J}_{q2}.$ Thus $j_{q2}$ can be
written as $j_{q2}=\partial_{\tilde x}^2\mathcal{J}_{q2}$ or
$j_{q2}=\partial_{\tilde x}\mathcal{J}_{q1}^{eff}.$ The currents
$j_{q2},$ $\mathcal{J}_{q1}^{eff}$ and $\mathcal{J}_{q2}$ are at
different levels: the basic current $\mathcal{J}_{q2}$ is at the
deepest level because it has not a substructure; $j_{q2}$ is at
the highest level because it has two substructures. The index $i$
in $j_{qi}$ is called "rank" $i$, i.e. the "rank" refers to the
number of the sub-currents included in $j_{qi}.$

In the $(\hat x,\hat p)$ system  from Eqs.~(\ref{Eq:hat-H}),
 (\ref{Eq:xi-Hun}) and  (\ref{Eq:H-q-summed}),
to the order $f^2,$ the corresponding perturbation $q$-deformed
current  $\hat j^{(2)}_q(\hat x, t)$ is
\begin{equation}
\label{Eq:jq-hat}
\hat j^{(2)}_q(\hat x, t)=i\frac{f^2}{6}
\Bigl \{\psi^{*}(\hat x,t) \hat x^3 V'(\hat x) \partial_{\hat x}
 \psi(\hat x,t)
-\bigl [ \partial_{\hat x} \psi^{*}(\hat x,t) \bigr ]
 \hat x^3 V'(\hat x)\psi(\hat x,t) \Bigr \} .
\end{equation}

Locally the structures of  $\hat j^{(2)}_q(\hat x, t)$ and
$\tilde j^{(2)}_q(\tilde x, t)$ are different.
The current $\hat j^{(2)}_q(\hat x, t)$ is explicitly potential dependent.
But the  potential is not explicitly included in the current
$\tilde j^{(2)}_q(\tilde x,t).$
Because the wave function $\psi (\tilde x, t)$ is  potential dependent,
the current $\tilde j^{(2)}_q(\tilde x,t)$ is implicitly  potential dependent.

A question arises:  whether the two  currents
 $\hat j^{(2)}_q(\hat x, t)$ and $\tilde j^{(2)}_q(\tilde x,t)$
globally lead to the same result in physics?
The question is answered by the following theorem.

{\bf Perturbation Equivalence Theorem of $q$-Deformed Currents}
The total contributions of the two
currents $\hat j^{(2)}_q(\hat x, t)$ and $\tilde j^{(2)}_q(\tilde x, t)$
 to the whole space are the same.

First we notice that for  the eigenstate $|\psi_n^{(0)}\rangle$ of the
undeformed Hamiltonian $H_{\rm un},$
 $H_{\rm un}|\psi_n^{(0)}\rangle= E^{\rm(un)}_n|\psi_n^{(0)}\rangle,$
it is natural to assume that the structure of the undeformed wave function
$\psi_n^{(0)}(\hat x_0)=\langle \hat x_0|\psi_n^{(0)}\rangle$
in the configuration space $\hat x_0$ and
the structure of the undeformed wave function
$\psi_n^{(0)}(\tilde x_0)=\langle \tilde x_0|\psi_n^{(0)}\rangle$
in the configuration space $\tilde x_0$ are the same.

Now the demonstration of the equivalence theorem is simple.
Integrating $\hat j^{(2)}_q(\hat x, t)$ and
$\tilde j^{(2)}_q(\tilde x, t)$, respectively, in configuration
spaces $\hat x_0$ and $\tilde x_0$ we obtain their total
contributions  to the whole space (for our purpose common
numerical factor is not important)
\begin{eqnarray}
\label{Eq:J-hat}
\hat J^{(2)}_q &=& \int d\hat x_0 \hat j^{(2)}_q(\hat x_0, t)
\nonumber \\
&\sim& \int d\hat x_0 \bigl[\psi_n^{(0)\ast}(\hat x_0 )
\hat x_0^3 V'\partial_{\hat x_0} \psi_n^{(0)}(\hat x_0 )-
(\partial_{\hat x_0}\psi_n^{(0)\ast}(\hat x_0))
\hat x_0^3 V' \psi_n^{(0)}(\hat x_0) \bigr]  \nonumber \\
&\sim& \int d\hat x_0 \psi_n^{(0)\ast}(\hat x_0 )
\bigl[2i \hat x_0^3 V'\hat p+3\hat x_0^2 V' +
\hat x_0^3 V'' \bigr]\psi_n^{(0)}(\hat x_0 )
\end{eqnarray}
and
\begin{eqnarray}
\label{Eq:J-tilde}
\tilde J^{(2)}_q &=& \int d\tilde x_0 \tilde j^{(2)}_q(\tilde x_0, t)
=\int d\tilde x_0 \bigl[-4\bigl(\tilde j^{(2)}_{q0}(\tilde x_0, t)
-\frac{1}{6} j^{(2)}_{un}(\tilde x_0, t)\bigr]  \nonumber \\
&\sim& \int d\tilde x_0 \bigl[ -4\bigl((\partial_{\tilde x_0}
\psi_n^{(0)\ast}(\tilde x_0 ))
\tilde x_0^2 \partial_{\tilde x_0}^2 \psi_n^{(0)}(\tilde x_0 )
-(\partial_{\tilde x_0}^2\psi_n^{(0)\ast}(\tilde x_0 ))
\tilde x_0^2 \partial_{\tilde x_0} \psi_n^{(0)}(\tilde x_0 )\bigr)
 \nonumber \\
&&-\bigl(\psi_n^{(0)\ast}(\tilde x_0 )\partial_{\tilde x_0}
\psi_n^{(0)}(\tilde x_0 )-
(\partial_{\tilde x_0}\psi_n^{(0)\ast}(\tilde x_0 ))
\psi_n^{(0)}(\tilde x_0 )\bigr)\bigr]
\nonumber \\
&\sim& \int d\tilde x_0 \psi_n^{(0)\ast}(\tilde x_0 )
\bigl[2i\tilde x_0^2 \tilde p^3+6\tilde x_0 p^2 -
2i\tilde p \bigr]\psi_n^{(0)}(\tilde x_0 )
\end{eqnarray}
In the undeformed stationary state $|\tilde \psi^{(0)}>$ we have
\begin{equation}
\label{Eq:Virial}
i\frac{d}{dt} \langle \tilde \psi^{(0)}|\tilde  x^m \tilde p^n
|\tilde \psi^{(0)}\rangle
= \langle \tilde\psi^{(0)}|\left[ \tilde  x^m\tilde  p^n,
\frac{1}{2\mu} \tilde  p^2+V(\tilde  x )\right]
|\tilde\psi^{(0)}\rangle=0.
\end{equation}
Using Eq.~(\ref{Eq:Virial}) and considering the cases of ($m=3,$ $n=2$),
($m=2,$ $n=1$),  and ($m=1,$ $n=0$), we have
\begin{equation}
\frac{1}{\mu}\langle \tilde\psi^{(0)}|\tilde  x^2 \tilde p^3
|\tilde\psi^{(0)}\rangle
=\frac{i}{\mu}\langle \tilde\psi^{(0)}|\tilde  x \tilde p^2
|\tilde\psi^{(0)}\rangle
+\frac{2}{3}\langle \tilde\psi^{(0)}|\tilde  x^3 V'\tilde p
|\tilde\psi^{(0)}\rangle
-\frac{i}{3}\langle \tilde\psi^{(0)}|\tilde  x^3V''
|\tilde\psi^{(0)}\rangle,
\nonumber
\end{equation}
\begin{equation}
\frac{1}{\mu}\langle \tilde\psi^{(0)}|\tilde  x \tilde p^2
|\tilde\psi^{(0)}\rangle
=\frac{1}{2}\langle \tilde\psi^{(0)}|\tilde  x^2 V'
|\tilde\psi^{(0)}\rangle
\nonumber
\end{equation}
\begin{equation}
\langle \tilde\psi^{(0)}|\tilde p|\tilde\psi^{(0)}\rangle=0
\nonumber
\end{equation}
Putting these three equations in Eq.~(\ref{Eq:J-tilde}), shows
that the integral of Eq.~(\ref{Eq:J-tilde}) is just the integral
of Eq.~(\ref{Eq:J-hat}).

Furthermore, it is clarified that the time evolution of the
$q$-deformed dynamics  and the perturbation shifts of the energy
spectrum  are the same in the $(\hat x,\hat p)$ system and the
$(\tilde x,\tilde p)$ system \cite{JZZ01}. We conclude that the
$(\hat x,\hat p)$ system and  the $(\tilde x,\tilde p)$ system are
equivalent in describing the $q$-deformed dynamics.

In the ordinary quantum theory, there is only a trivial
transformation of canonical variables $(x, p)$ which remains
commutation relations. But unlike the ordinary quantum theory, in
the $q$-deformed quantum theory there is a  non-trivial
transformation  among two pairs of the undeformed variables which
remains the commutation relations. The operator  $F^{-1}(\hat z)$
defined by Eq.~(\ref{Eq:F(z)})
 is non-unitary.
It is a variable transformation between  two undeformed variables
$(\hat x,\hat p)$ and $(\tilde x,\tilde p)$;
it should be distinguished from a unitary transformation in a
Hilbert space.
Thus it is not clear whether two perturbation formulations in
the $(\hat x,\hat p)$ system and the $(\tilde x,\tilde p)$ system
are equivalent.
It is interesting to clarify that though locally the structures
of two $q$-deformed  perturbation probability currents
are different,
but their total contributions  to the whole space are the same.

The structures of the perturbation $q$-deformed currents  show
that the essential deviation of the $q$-deformed quantum mechanics
from the ordinary quantum mechanics is in their {\it local}
structure. Further exploration of novel local properties in the
$q$-deformed quantum mechanics is promising.

\vspace{0.4cm}
  This work has been supported by the Deutsche Forschungsgemeinschaft
(Germany).
 The author would like to thank  W. R\"uhl for helpful discussions.
 His work has also been supported by the National Natural Science
Foundation of China under the grant number 10074014 and by the Shanghai
Education Development Foundation.


\end{document}